# Plasma and cavitation dynamics during pulsed laser microsurgery *in vivo*

M. Shane Hutson[*] and Xiaoyan Ma

*Department of Physics & Astronomy and Vanderbilt Institute for Integrative Biosystem Research & Education, VU Station B #351807, Vanderbilt University, Nashville, TN 37235-1807*

We compare the plasma and cavitation dynamics underlying pulsed laser microsurgery in water and in fruit fly embryos (*in vivo*) – specifically for nanosecond pulses at 355 and 532 nm. We find two key differences. First, the plasma-formation thresholds are lower *in vivo* – especially at 355 nm – due to the presence of endogenous chromophores that serve as additional sources for plasma seed electrons. Second, the biological matrix constrains the growth of laser-induced cavitation bubbles. Both effects reduce the disrupted region *in vivo* when compared to extrapolations from measurements in water.

  52.38.Mf    Laser ablation
  52.50.Jm    Plasma production and heating by laser beams
  47.55.dp    Cavitation and boiling

    Pulsed laser microsurgery has emerged as an important technique for probing biological systems through the targeted disruption of cellular and sub-cellular structures [1, 2]. Applications include cell lineage studies in developing organisms [3], sampling of heterogeneous systems via laser pressure catapulting [4], and gene delivery through transient membrane disruption (optoporation) [5]. Additionally, new applications have focused on the dissection of cytoskeletal filaments – either as probes of intracellular forces in adherent cells [6] or of intercellular forces in developing embryos [7]. Even with an abundance of microsurgical applications, there have been just a few attempts to characterize the underlying physical mechanisms [8, 9]. Most recently, Venugopalan *et al* [10] provided crucial clues that pulsed laser microsurgery is driven by laser-induced plasma formation. The expanding plasma subsequently drives shock wave propagation as well as the dynamic expansion and collapse of cavitation bubbles. One of the most puzzling findings from these studies was that the cavitation bubbles had radii of 45-470 μm. Under optimal conditions in living tissues, the scale of the laser-disrupted region is just a few hundred nanometers [2].

    The above conflict points to a limitation in our physical understanding of pulsed laser microsurgery. The initial studies were conducted only in distilled water – and only at visible and near-infrared wavelengths. No one has systematically explored the impact of the biological matrix or the role of endogenous chromophores. Such chromophores may play a large role in microsurgery at near-UV wavelengths. We address these issues

by comparing the plasma and cavitation dynamics during laser microsurgery in water to that in living fruit fly embryos (*in vivo*) – both at visible and near-UV wavelengths.

We focus either the 2$^{nd}$ or 3$^{rd}$ harmonic (532 nm or 355 nm) of a *Q*-switched Nd:YAG laser (4 ns pulsewidth) through the 40×, 1.3 NA oil-immersion objective of a Zeiss LSM410 inverted confocal microscope. The entire laser-microsurgery system has been described in detail previously [11] and allows us to take fluorescent images of thick biological samples while simultaneously cutting the samples with single (or multiple) pulses at any user-defined location (or trajectory). The transgenic fruit flies used in these experiments produce a green fluorescent protein (GFP):E-cadherin chimera that fluorescently labels the epithelial cell borders [12]. The fruit fly embryos are approximately ellipsoidal (~500 by ~200 μm) and are arranged on a coverslip with their long axis parallel to the surface. The fly embryos are covered with distilled water and a needle hydrophone (HNR-0500, Onda Corporation, 0.5 mm aperture, <20 ns rise time) is placed 1.5 mm above the targeted fruit fly embryo to record the pressure transients associated with plasma formation, i.e. shock wave propagation and cavitation bubble collapse. The time between these two transients is a direct measure of the cavitation bubble oscillation time, $T_{osc}$.

Figure 1 shows how strongly the *in vivo* effects of laser microsurgery depend on wavelength. At 532 nm, with an energy just above the plasma-formation threshold, our attempt to cut a single cell edge in a fly embryo destroyed all cellular structure over a wide area (>80 μm in diameter). At 355 nm, and 5× threshold, we disrupted just the targeted cell edge. The two cells sharing this edge then expand over tens of seconds as the fly epithelium comes to a new mechanical equilibrium. Strong clues as to the root of these differences come from simultaneous hydrophone measurements. The measured $T_{osc}$ can be used to calculate the maximum bubble radius ($R_{max}$) after Rayleigh [13]. Although this equation is not strictly valid *in vivo* (a viscoelastic medium and lacks spherical symmetry), our direct observations (below) show that a predicted $R_{max}$ is good to within ±15%. In these two examples, $R_{max}$ was 5× larger in the 532-nm ablation – matching the much larger disrupted region.

The differences in $R_{max}$ above are a direct consequence of large differences in the plasma-formation threshold *in vivo*. Although this threshold is typically measured by observing visible flashes from plasma luminescence, this method does not work well for *in vivo* samples at 355 nm due to very strong fluorescence from endogenous chromophores. Thus, we quantified the plasma thresholds by observing the presence or absence of shock waves in the hydrophone data (and fitting the resulting probabilities to a Gaussian error function [14]). For 532-nm ablation, we confirmed that the two methods gave identical results. In distilled water, the plasma formation thresholds at 355 and 532 nm were 1.86 and 13.28 μJ, respectively. This 532-nm threshold is considerably higher than previously reported [10] due to known sources of spherical aberration in our beamline [15]. Nonetheless, the 7-fold increase in threshold at the longer wavelength is consistent with previous comparisons of 532 nm and 1064 nm where a nearly 10-fold increase was observed [10]. Surprisingly, we find even larger differences *in vivo*, where the thresholds at 355 and 532 nm were 0.23 and 8.63 μJ, respectively – a 38-fold increase at the longer wavelength. This large difference is mainly due to the fact that the *in vivo* threshold at 355 nm is just 1/8$^{th}$ of the threshold in water.

Once above threshold, the subsequent cavitation dynamics are not strongly wavelength dependent. Measurements of $T_{osc}$ (and $R_{max}$) for a wide range of pulse energies at both 355 and 532 nm are shown in Figure 2. The most striking characteristic of this plot is that all of the measurements in water (filled symbols) fall along a single curve. This includes data previously reported by others for 532 and 1064 nm [10]. Similarly, almost all of the measurements *in vivo* (open symbols) fall along a different single curve – one with smaller bubble radii. The few exceptions are for high energy ablations at 532 nm conducted after the vitelline membrane encasing a fly embryo was ruptured by an earlier pulse. These points fall along the curve describing cavitation dynamics in water, suggesting that the main difference between the "in water" and *in vivo* curves is mechanical constraint of the cavitation bubbles. In accord with this explanation, the *in vivo* cavitation bubbles plateau at high energies with an $R_{max}$ just below 70 μm, but the bubbles in water continue to grow. The smallest cavitation bubbles observed here are those *in vivo* at 355 nm with an $R_{max}$ of just 3.2 μm.

The wavelength-dependent thresholds and the nearly wavelength-independent cavitation dynamics above threshold correlate well with models of plasma formation during nanosecond pulses [16]. The plasma-formation threshold is largely determined by the intensity needed to produce quasi-free seed electrons – by mulitphoton ionization in water. Shorter wavelengths can drive multiphoton ionization at lower intensities and thus have lower thresholds. In contrast, the cavitation dynamics are largely determined by the energy content of the plasma. This final energy content is largely determined by the efficiency with which seed electrons drive formation of the full plasma through cascade ionization. This process is much more dependent on the laser pulse energy than on the wavelength.

What then accounts for the greatly reduced plasma formation threshold at 355 nm *in vivo*? The models suggest we should look for a new source of seed electrons. Our primary candidate is the reduced form of nicotinamide adenine dinucleotide (NADH). This ubiquitous biomolecule is present in relatively high (mM) concentrations in developing embryos [17]; it has a strong and broad absorption maximum at 340 nm [18]; and it has been shown to undergo one-electron oxidation when irradiated at 353 nm by a sequential two-photon process [19]. As opposed to a multiphoton process, this sequential two-photon ionization could generate seed electrons at much lower intensities. To test this hypothesis, we irradiated aqueous solutions of NADH at 355 nm (buffered with 10 mM $CH_3COOH$, pH 5.0). We find that the plasma-formation threshold can be reduced to what we observed *in vivo* at an NADH concentration of 38 mM. This is higher than the average physiological range; however, NADH is not homogeneously distributed in cells, but is instead concentrated in mitochondria. As for the laser-induced cavitation dynamics, $T_{osc}$ (and $R_{max}$) are larger in NADH solution than *in vivo* – additional evidence that the biological matrix constrains the cavitation bubbles. As shown in Figure 3, the curve of $T_{osc}$ versus pulse energy in NADH solution simply extends the in water curve to lower pulse energies.

In the discussions above, we have repeatedly used the Rayleigh formula to calculate $R_{max}$ for each measured $T_{osc}$. To confirm the validity of this formula under our non-ideal conditions, we have directly imaged passage of the bubble front at various distances from the ablation site. To do so, we focus the microscope's 647 nm Ar-Kr line to a specific location in the focal plane, and collect backscattered light through a confocal pinhole to a photomultiplier tube. Both in water and *in vivo*, very little light is normally backscattered. The ablating laser is then targeted to a different spot in the focal plane a distance $r$ from the imaging laser. If the cavitation bubble expands far enough to reach the imaging laser, then the amount of backscattered light increases dramatically as the bubble front passes by (at time $t_1$), and remains high until the bubble front passes by again during bubble collapse (at time $t_2$). From each backscattered signal trace, we get two points in the curve of bubble radius versus time, i.e. $r(t_1)$ and $r(t_2)$. By changing the distance between the imaging and ablating lasers, we trace out entire $r(t)$ curves. Since $T_{osc}$ (and $R_{max}$) vary from pulse to pulse, we normalize our times by $T_{osc}$ and our distances by $R_{max}$, as shown in Figure 4. In distilled water, this procedure traces out a very smooth trajectory with the largest cavitation bubbles reaching radii of $0.86 R_{max}$ and with no evidence for bubble passage above $0.93 R_{max}$. Thus the cavitation bubbles appear to be 7-14% smaller than predicted. The i*n vivo* bubble trajectory is not as reproducible. Here we find bubbles reaching radii of $1.16 R_{max}$; and there are examples with no evidence for bubble passage as small as $1.11 R_{max}$. Thus the *in vivo* cavitation bubbles appear to be 11-16% larger than predicted. The in water versus *in vivo* differences likely result from the different ways spherical symmetry is broken. Bubbles in water can expand freely away from the coverslip and out of the focal plane. Bubbles *in vivo* are constrained between embryonic tissue layers and can only expand freely within the focal plane.

After confirming that the Rayleigh formula is a reasonable approximation *in vivo*, we can use our hydrophone data to estimate the energy content of both the cavitation bubble and the shock wave [10]. These estimates are plotted in Figure 5 as fractions of the laser pulse energy. Just as in the plot of $T_{osc}$, the plot of cavitation bubble energy shows one wavelength-independent curve in water and a different wavelength-independent curve *in vivo*. In water, the cavitation bubbles may retain up to 20% of the laser pulse energy, but this maximum is just 1% *in vivo*. The remaining energy is taken up by deformation of the surrounding biological matrix. In the plot of shock wave energy, the distinction between in water and *in vivo* data is lost. All of the data fall along a single trend line that plateaus with shock waves representing 15% of the laser pulse energy. Note that this estimate is for a distance of 1.5 mm from the ablation site and is a lower bound on the shock wave energy at the ablation site. At very low energies, where ablation is only possible at 355 nm and *in vivo*, the cavitation bubble and shock wave both retain less than 0.01% of the laser pulse energy.

In conclusion, these experiments have shown that the size of laser-induced cavitation bubbles is a major determinant of the region disrupted during *in vivo* laser microsurgery. Compared to measurements in water, the cavitation bubbles induced *in vivo* are much smaller due to two effects: (1) the plasma-formation threshold can be greatly reduced by the presence of endogenous chromophores like NADH that serve as sources of seed electrons at lower laser intensities; and (2) the expansion of cavitation bubbles is constrained by the surrounding matrix. In 355-nm laser microsurgery on fruit fly embryos, the former effect dominates (roughly by a factor of two). In other applications, the relative importance may be reversed – for example, when the surrounding matrix is stiffer, or when there is a lack of endogenous seed electron sources, or when the laser pulse is so short that seed electron generation is still dominated by multiphoton ionization of water. In any case, optimization of pulsed laser microsurgery cannot rely solely on extrapolations from water, but must carefully consider both *in vivo* effects.

ACKNOWLEDGEMENTS

This work supported by VIIBRE (Vanderbilt Institute for Integrative Biosystem Research & Education) and the National Science Foundation CAREER Program (IOB-0545679).

---

[*]Corresponding author: shane.hutson@vanderbilt.edu.

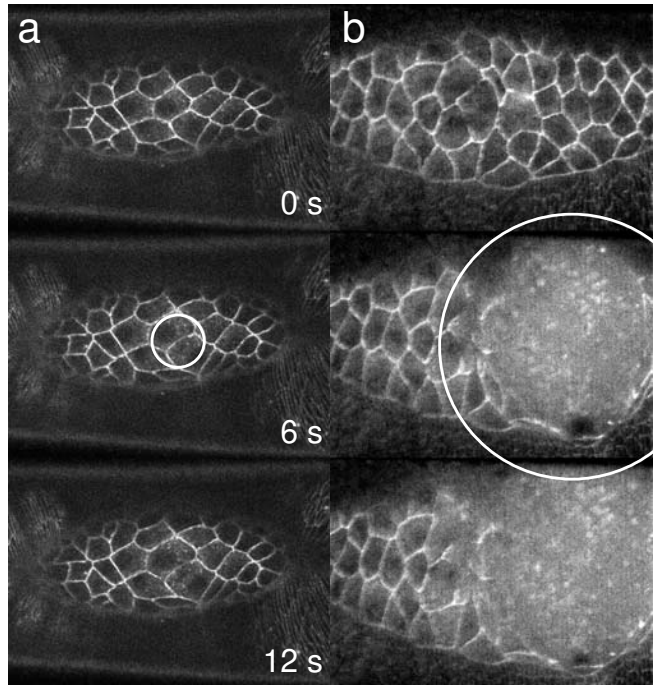

FIG. 1. Assessment of cavitation damage *in vivo*. Three sequential confocal images are shown of a fruit fly embryo before and after ablation at (a) 5× threshold and $\lambda = 355$ nm or (b) 1× threshold and $\lambda = 532$ nm. Ablation occurred between the first and second panels of each sequence. The white circle in the second panel is centered on the ablation site and has a radius equal to the calculated $R_{max} = 12.9$ μm in (a) and 65.6 μm in (b).

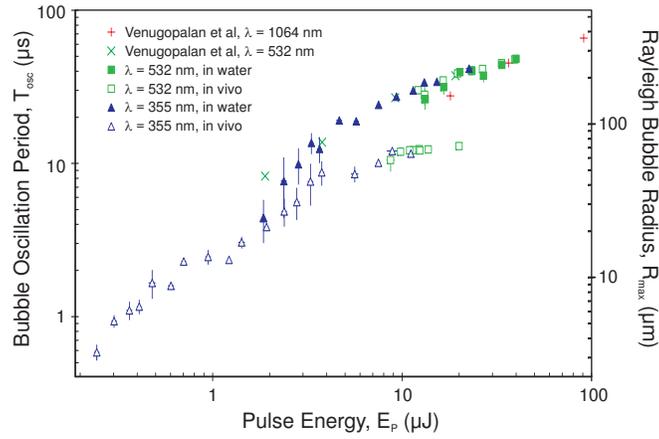

FIG. 2. Cavitation bubble parameters versus laser pulse energy in water and *in vivo*. To facilitate comparison, we have included data previously reported for ablation in water at 532 and 1064 nm [10].

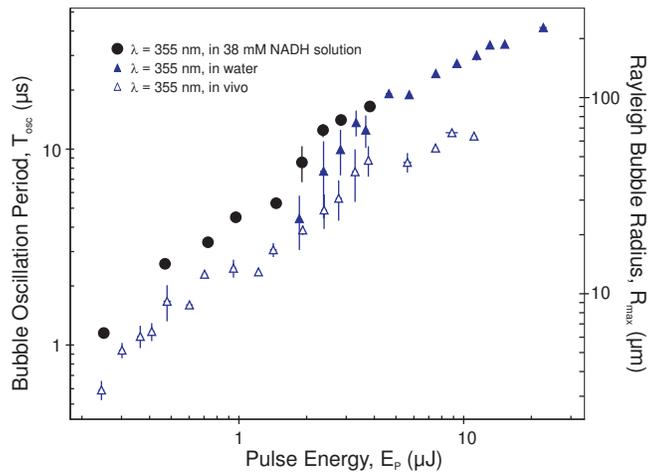

FIG. 3. Cavitation bubble parameters from 355-nm ablation in a 38 mM NADH solution. For comparison, data from 355-nm ablation in water and *in vivo* are repeated here.

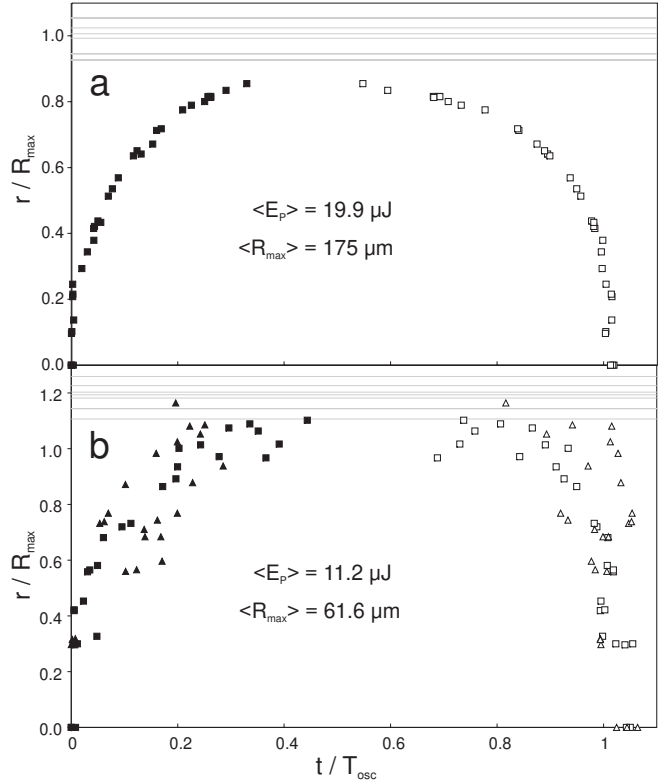

FIG. 4. Direct measurements of the growth and collapse of ablation-induced cavitation bubbles via confocal laser backscatter: (a) in water at 532 nm; and (b) *in vivo* at 532 nm. The average pulse energy, $E_P$, and calculated $R_{max}$ for each data set are given. Filled symbols represent passage of the bubble front during expansion; open symbols represent the passage during bubble collapse. The grey lines represent distances at which we saw no evidence of bubble passage. For *in vivo* experiments, the separation between the ablation site and the imaging laser is parallel to the long axis (■) or short axis (▲) of the embryo.

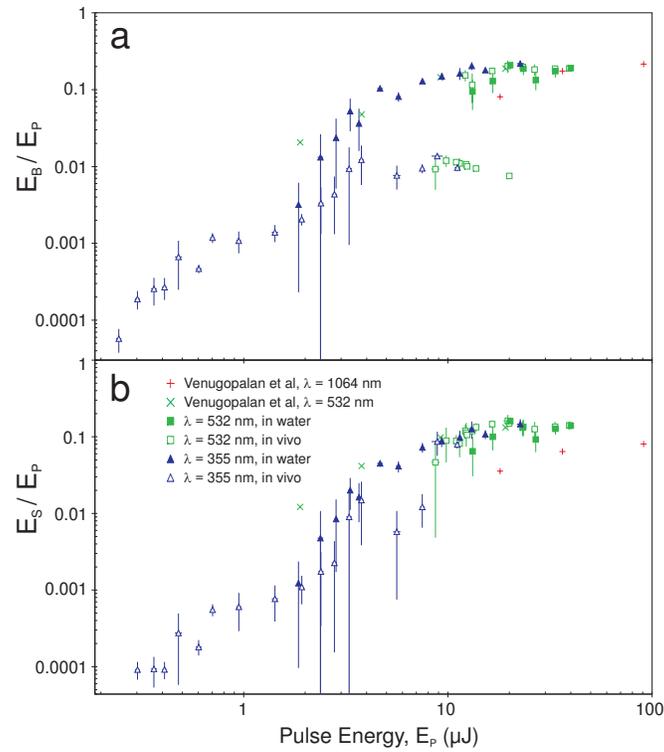

FIG. 5. Fraction of the laser pulse energy represented in (a) the maximally expanded cavitation bubble and (b) the shock wave at 1.5 mm from the ablation site. To facilitate comparison, we have included data previously reported for ablation in water at 532 and 1064 nm [10].